\documentclass[smallextended]{svjour3}

\usepackage{graphicx,amsmath,amssymb,amsfonts}
\usepackage[margin=1in]{geometry}
\usepackage{hyperref}
\usepackage{natbib}
\usepackage{color}

\graphicspath{{../correlation_analysis/figures/}}

\makeatletter
\newcommand{\customlabel}[2]{%
\protected@write \@auxout {}{\string \newlabel {#1}{{#2}{\thepage}{}{}{}}}}
\makeatother

\customlabel{fig:cor_entropy_PC_screen_CSrmsf}{S1}
\customlabel{fig:cor_omega_PC_screen}{S2}
\customlabel{fig:cor_omega_PC_screen_CSrmsf}{S3}

\journalname{Preprint}

\begin{document}

\title{Predicting evolutionary site variability from structure in viral proteins: buriedness, packing, flexibility, and design}
\titlerunning{Predicting evolutionary site variability from structure}

\author{Amir Shahmoradi \and Dariya K. Sydykova \and \\ Stephanie J. Spielman \and Eleisha L. Jackson \and \\ Eric T. Dawson \and Austin G. Meyer \and Claus O. Wilke}

\institute{Amir Shahmoradi \at Department of Physics, The University of Texas at Austin, TX, 78712. \\
\and
Amir Shahmoradi \and Dariya K. Sydykova \and Stephanie J. Spielman \and Eleisha L. Jackson \and Eric T. Dawson \and Austin G. Meyer \and Claus O. Wilke \at Department of Integrative Biology, Center for Computational Biology and Bioinformatics, and Institute for Cellular and Molecular Biology, The University of Texas at Austin, TX, 78712. \email{wilke@austin.utexas.edu} }

\date{}

\maketitle

\begin{abstract}
Several recent works have shown that protein structure can predict site-specific evolutionary sequence variation. In particular, sites that are buried and/or have many contacts with other sites in a structure have been shown to evolve more slowly, on average, than surface sites with few contacts. Here, we present a comprehensive study of the extent to which numerous structural properties can predict sequence variation. The quantities we considered include buriedness (as measured by relative solvent accessibility), packing density (as measured by contact number), structural flexibility (as measured by B factors, root-mean-square fluctuations, and variation in dihedral angles), and variability in designed structures. We obtained structural flexibility measures both from molecular dynamics simulations performed on 9 non-homologous viral protein structures and from variation in homologous variants of those proteins, where available. We obtained measures of variability in designed structures from flexible-backbone design in the Rosetta software. We found that most of the structural properties correlate with site variation in the majority of structures, though the correlations are generally weak (correlation coefficients of 0.1 to 0.4). Moreover, we found that buriedness and packing density were better predictors of evolutionary variation than was structural flexibility. Finally, variability in designed structures was a weaker predictor of evolutionary variability than was buriedness or packing density, but it was comparable in its predictive power to the best structural flexibility measures. We conclude that simple measures of buriedness and packing density are better predictors of evolutionary variation than are more complicated predictors obtained from dynamic simulations, ensembles of homologous structures, or computational protein design.
\end{abstract}

\section*{Introduction}

Patterns of amino-acid sequence variation in protein-coding genes are shaped by the structure and function of the expressed proteins
\citep{WilkeDrummond2010,Liberlesetal2012,MarshTeichmann2014}. As the most basic reflection of this relationship, buried residues in proteins tend to be more evolutionarily conserved than exposed residues
\citep{Overingtonetal1992,Goldmanetal1998,MirnyShakhnovich1999,Deanetal2002}. More specifically, when evolutionary variation is plotted as a function of Relative Solvent Accessibility (RSA, a measure of residue buriedness), the relationship falls, on average, onto a straight line with a positive slope \citep{FranzosaXia2009,Ramseyetal2011,FranzosaXia2012,Scherreretal2012}. Importantly, however, this relationship represents an average over many sites and many proteins. At the level of individual sites in individual proteins, RSA is often only weakly correlated with evolutionary variation \citep{MeyerWilke2013,Meyeretal2013,Yehetal2014}.

Other structural measures, such as residue contact number (CN), have also been shown to correlate with sequence variability \citep{Liaoetal2005,FranzosaXia2009,Yehetal2014}, and some have argued that CN predicts evolutionary variation better than RSA \citep{Yehetal2014,Yehetal2014b}. Because CN may be a proxy for residue and site-specific backbone flexibility \citep{Halle2002}, a positive trend between local structural variability and sequence variability may also exist \citep{Yehetal2014}. Indeed, several authors have suggested that such protein dynamics may play a role in sequence variability \citep{LiuBahar2012,NevinGereketal2013,MarshTeichmann2014}. However, a recent paper argued against the flexibility model, on the grounds that evolutionary rate is not linearly related to flexibility \citep{Huangetal2014}.

While RSA and CN can be calculated in a straightforward manner from individual crystal structures, measures of structural flexibility, either at the side-chain or the backbone level, are more difficult to obtain. Two viable approaches to measuring structural flexibility are (i) examining existing structural data or (ii) simulating protein dynamics. NMR ensembles may approximate physiologically relevant structural fluctuations. Similar fluctuations are observed in ensembles of homologous crystal structures \citep{Maguidaetal2008,EchaveFernandez2010}. The thermal motion of atoms in a crystal are recorded in B factors, which are available for every atom in every crystal structure. To measure protein fluctuations using a simulation approach, one can either use coarse-grained modeling, e.g. via Elastic Network Models \citep{Sanejouand2013}, or atom-level modeling, e.g. via molecular dynamics (MD) \citep{KarplusMcCammon2002}. However, it is not well understood which, if any, of these measures of structural flexibility provide insight into the evolutionary process, in particular into residue-specific evolutionary variation.

Here, we provide a comprehensive analysis of the extent to which numerous different structural quantities predict evolutionary sequence (amino-acid) variation. We considered two measures of evolutionary sequence variation: site entropy, as calculated from homologous protein alignments, and evolutionary rate. As structural predictors, we included buriedness (RSA), packing density (CN), and measures of structural flexibility, including B factors, several measures of backbone and side-chain variability obtained from MD simulations, and backbone variability obtained from alignments of homologous crystal structures. We additionally considered site variability, as predicted from computational protein design with Rosetta. 

On a set of nine viral proteins, RSA and CN generally performed better at predicting evolutionary site variation than did either measures of structural flexibility or computational protein design. Among the measures of structural flexibility, measures of side-chain variability performed better than do measures of backbone variability, possibly because the former are more tightly correlated with residue packing. Finally, site variability predicted from computational protein design performed worse than the best-performing measures of structural fluctuations.

\section*{Materials and Methods}

\subsection*{Sequence data, alignments, and evolutionary rates}

All viral sequences except influenza sequences were retrieved from \url{http://hfv.lanl.gov/components/sequence/HCV/search/searchi.html}.
The sequences were truncated to the desired genomic region but not in any other way restricted. Influenza sequences were downloaded from \url{http://www.fludb.org/brc/home.spg?decorator=influenza}. We only considered human influenza A, H1N1, excluding H1N1 sequences derived from the 2009 Swine Flu outbreak or any sequence from before 1998, but we did not place any geographic restrictions.

For all viral sequences, we removed any sequence that was not in reading frame, any sequence which was shorter than 80\% of the longest sequence for a given viral protein (so as to remove all partial sequences), and any sequence containing any ambiguous characters. Alignments were constructed using amino-acid sequences with MAFFT \citep{Katohetal2002,Katohetal2005}, specifying the \verb+--auto+ flag to select the optimal algorithm for the given data set, and then back-translated to a codon alignment using the original nucleotide sequence data.

To assess site-specific sequence variability in amino-acid alignments, we calculated the Shannon entropy ($H_i$) at each alignment column $i$:
\begin{equation}
        H_i = - \sum_jP_{ij}\ln P_{ij},
\end{equation}
where $P_{ij}$ is relative frequency of amino acid $j$ at position $i$ in the alignment.

For each alignment, we also calculated evolutionary rates, as described \citep{SpielmanWilke2013}. In brief, we generated a phylogeny for each codon alignment in RAxML \citep{RaxMLHPC} using the GTRGAMMA model. Using the codon alignment and phylogeny, we inferred evolutionary rates with a Random Effects Likelihood (REL) model, using the HyPhy software \citep{KosakovskyPondetal2005}. The REL model was a variant of the GY94 evolutionary model \citep{GoldmanYang1994} with five $\omega$ rate categories as free parameters.  We employed an Empirical Bayes approach \citep{Yang2000} to infer $\omega$ values for each position in the alignment. These $\omega$ values represent the evolutionary-rate ratio $dN/dS$ at each site.

\subsection*{Protein crystal structures}

A total of 9 viral protein structures were selected for analysis, as tabulated in Table \ref{tab:pdb_names}. Sites in the PDB structures were mapped to sites in the viral sequence alignments via a custom-built python script that creates a consensus map between a PDB sequence and all sequences in an alignment.

For each of the viral proteins, homologous structures were identified using the \texttt{blast.pdb} function of the R package Bio3D \citep{Grantetal2006}. BLAST hits were retained if they had $\geq35$\% sequence identity and $\geq90$\% alignment length. Among the retained hits, we subsequently identified sets of homologous structures with unique sequences and with mutual pairwise sequence divergences of $\geq2\%$, $\geq5\%$, and $\geq10\%$.

\subsection*{Molecular Dynamics Simulations}

Molecular dynamics (MD) simulations were carried out using the GPU implementation of the {\it Amber12} simulation package \citep{SalomonFerreretal2013} with the most recent release of the Amber fixed-charge force field (ff12SB; c.f., AmberTools13 Manual). Prior to MD production runs, all PDB structures were first solvated in a box of TIP3P water molecules \citep{jorgensen1983} such that the structures were  at least $10\AA$ away from the box walls. Each individual system was then energy minimized using the steepest descent method for 1000 steps, followed by conjugate gradient for another 1000 steps. Then, the structures were constantly heated from 0K to 300K for 0.1ns, followed by 0.1ns constant pressure simulations with positional harmonic restraints on all atoms to avoid instabilities during the equilibration process. The systems were then equilibrated for another 5ns without positional restraints, each followed by 15ns of production simulations for subsequent post-processing and analyses. All equilibration and production simulations were run using the SHAKE algorithm \citep{Ryckaert1977}. Langevin dynamics were used for temperature control.

\subsection*{Measures of buriedness, packing density, and structural flexibility}

As a measure of residue buriedness, we calculated Relative Solvent Accessibility (RSA). To calculate RSA, we first calculated the Accessible Surface Area (ASA) for each residue in each protein, using the DSSP software \citep{KabschSander1983}. We then normalized ASA values by the theoretical maximum ASA of each residue \citep{Tienetal2013} to obtain RSA. We considered two measures of local packing density, contact number (CN) and weighted contact number (WCN). We calculated CN for each residue as the total number of C$\alpha$ atoms surrounding the  C$\alpha$ atom of the focal residue within a spherical neighborhood of a predefined radius $r_0$. Following \citet{Yehetal2014}, we used $r_0=13\AA$. We calculated WCN as the total number of surrounding C$\alpha$ atoms for each focal residue, weighted by the inverse square separation between the C$\alpha$ atoms of the focal residue and the contacting residue, respectively \citep{Shihetal2012}.

In most analyses, we actually used the inverse of CN and/or WCN, $\text{iCN}=1/\text{CN}$ and $\text{iWCN}=1/\text{WCN}$. Note that for Spearman correlations, which we use throughout here, replacing a variable by its inverse changes the sign of the correlation coefficient but not the magnitude.

As measures of structural flexibility, we considered RMSF, variability in backbone and side-chain dihedral angles, and B factors. We calculated RMSF for C$\alpha$ atoms based on both MD trajectories and homologous crystal structures. For MD trajectories, we calculated RMSF as
\begin{equation}
    \text{RMSF}_j = \Big[\sum_i \big(\mathbf{r}_i^{(j)}-\mathbf{r}_0^{(j)}\big)^2\Big]^{1/2}
\end{equation}
where $\text{RMSF}_j$ is the root-mean-square fluctuation at site $j$, $\mathbf{r}_i^{(j)}$ is the position of the C$\alpha$ atom of residue $j$ at MD frame $i$, and $\mathbf{r}_0^{(j)}$ is the position of the C$\alpha$ atom of residue $j$ in the original crystal structure.

To calculate RMSF from homologous structures, we first aligned the structures using the Bio3D package \citep{Grantetal2006}, and then we calculated
\begin{equation}
    \text{RMSF}_j = \Big[\sum_i w_i\big(\mathbf{r}_i^{(j)}-\langle\mathbf{r}^{(j)}\rangle\big)^2\Big]^{1/2},
\end{equation}
where $\mathbf{r}_i^{(j)}$ now stands for the position of the C$\alpha$ atom of residue $j$ in structure $i$, $\langle\mathbf{r}^{(j)}\rangle$ is the mean position of that C$\alpha$ atom over all aligned structures, and $w_i$ is a weight to correct for potential phylogenetic relationship among the aligned structures. The weights $w_i$ were calculated using BranchManager \citep{StoneSidow2007}, based on phylogenies built with RAxML as before.

To assess variability in backbone and side-chain dihedral angles, we calculated Var($\phi$), Var($\psi$), and Var($\chi_1$). The variance of a dihedral angle was defined according to the most common definition in directional statistics:  
First, a unit vector $\mathbf{x}_i$ is assigned to each dihedral angle $\alpha_i$ in the sample. The unit vector is defined as $\mathbf{x}_i = ( \cos (\alpha_i), \sin (\alpha_i) )$.
The variance of the dihedral angle is then defined as
\begin{equation}
\text{Var}(\alpha) = 1 - ||\langle \mathbf{x}\rangle||,
\end{equation}
where $||\langle \mathbf{x}\rangle||$ represents the length of the mean $\langle \mathbf{x}\rangle$, calculated as $\langle \mathbf{x}\rangle=\sum_i \mathbf{x}_i/n$. Here, $n$ is the sample size. The variance of a dihedral angle is, by definition, a real number in the range $[0,1]$, with $\text{Var}(\alpha) = 0$ corresponding to the minimum variability of the dihedral angle and $\text{Var}(\alpha) = 1$ to the maximum, respectively \citep{Berens2009}. Since the $\chi_1$ angle is undefined for Ala and Gly we excluded all sites with these residues in analyses involving $\chi_1$.

B factors were extracted from the crystal structures. We only considered the B factors of the C$\alpha$ atom of each residue.

\subsection*{Sequence Entropy from Designed Proteins}

Designed entropy was calculated as described \citep{Jacksonetal2013}. In brief, proteins were designed using RosettaDesign (Version 39284) \citep{LeaverFayetal2011} using a flexible backbone approach. This was done for all PDB structures in Table \ref{tab:pdb_names} as initial template structures. For each template, we created a backbone ensemble using the Backrub method \citep{Smith2008}. The temperature parameter in Backrub was set to 0.6, allowing for an intermediate amount of flexibility. We had previously found in a different data set that intermediate flexibility gave the highest congruence between designed and observed site variability \citep{Jacksonetal2013}.

For each of the 9 template structures we designed 500 proteins.

\subsection*{Availability of data and methods}

All details of simulations, input$/$output files, and scripts for subsequent analyses are available to view or download at \url{https://github.com/clauswilke/structural\_prediction\_of\_ER}.

\section*{Results}

\subsection*{Data set and structural variables considered}

Our goal in this work was to determine which structural properties best predict amino-acid variability at individual sites in viral proteins. To this end, we selected 9 viral proteins for which we had both high-quality crystal structures and abundant sequences to assess evolutionary sequence variation (Table~\ref{tab:pdb_names}). We quantified evolutionary variability in two ways: by calculating sequence entropies for each alignment column, and by calculating site-specific evolutionary-rate ratios $\omega=dN/dS$ (see Methods for details). Throughout this paper, we primarily report results obtained for sequence entropy. Results for $\omega$ were largely comparable, with some specific caveats detailed below.

As predictors of evolutionary variability, we considered buriedness, packing density, and residue flexibility. We additionally considered the variation seen in computationally designed protein variants. Buriedness quantifies the extent to which a residue is protected from solvent. We determined residue buriedness by calculating the relative solvent accessibility (RSA), which represents the relative proportion of a residue's surface in contact with solvent.

Packing density quantifies how many other residues a given residue interacts with. We determined packing density by calculating contact number (CN) and weighted contact number (WCN). CN counts the number of contacts within a sphere of a given radius around the $\alpha$-carbon of the focal residue, while WCN weights contacts by the distance between the two residues. Residue buriedness and packing density tend to be (anti-)correlated but measure qualitatively different properties of a residue. In particular, in the core of a protein, buriedness is always zero but packing density can vary. Because contact numbers decline as relative solvent accessibility increases, we replaced CN and WCN with their inverses, $\text{iCN}=1/\text{CN}$ and $\text{iWCN}=1/\text{WCN}$, in most analyses. Importantly, as Spearman rank correlations were used, this substitution only changed the sign of correlations but not the magnitude.

Measures of structural flexibility assess the extent to which a residue fluctuates in space as a protein undergoes thermodynamic fluctuations in solution. We quantified these fluctuations using several different measures. We considered B factors, which measure the spatial localization of individual atoms in a protein crystal, RMSF, the root mean-square fluctuation of the C$\alpha$ atom over time, and variability in side-chain and backbone dihedral angles, including Var($\chi_1$), Var($\phi$), and Var($\psi$). We employed two broad approaches, one using PDB crystal structures and one using molecular dynamics (MD) simulations, to obtain these measurements. Crystal structures yielded measures for B factors and RMSF; we obtained B factors from individual protein crystal structures, given in Table~\ref{tab:pdb_names}, and we calculated RMSF from aligned homologous crystal structures for those proteins which had sufficient sequence variation among crystal structures (see Methods and Table~\ref{tab:homologs} for details). MD simulations yielded measures for RMSF and variability in residue dihedral and side-chain angles. More specifically, we simulated MD trajectories for all crystal structures in Table~\ref{tab:pdb_names}. For each protein, we equilibrated the structure, simulated 15ns of chemical time, and recorded snapshots of the simulated structure every 10ps (see Methods for details). We obtained RMSF and angle variabilities from these snapshots. Additionally, we calculated time-averaged values of RSA, CN, and WCN. We also refer to these time-averaged measures as MD RSA, MD CN, and MD WCN, respectively. Unless specified otherwise, all results reported below were obtained using MD RSA, MD CN, and MD WCN.

As an alternative to predicting evolutionary variation from simple structural measures such as contact density or backbone flexibility, one can also predict evolutionary variation via a protein-design approach \citep{DokholyanShakhnovich2001,OllikainenKortemme2013,Jacksonetal2013}. In this case, one takes the protein structure of interest, replaces all residue side chains with randomly-chosen alternatives, and uses a coarse-grained or atom-level energy function to assess which side-chain choices are consistent with the backbone conformation of the focal structure. We have recently used this approach to compare natural and designed sequence variability in cellular proteins  \citep{Jacksonetal2013}, and we have found that (i) flexible-backbone design, where small backbone movements are allowed during the design phase, outperformed fixed-backbone design, and (ii) intermediate backbone flexibility, obtained via an intermediate design temperature, produced the highest congruence between designed and natural sequences. Similarly, \citet{DokholyanShakhnovich2001} had previously found that an intermediate temperature parameter gave the best agreement between designed and natural sequences in their model. Inspired by these prior results, we investigated here how protein design performed relative to simpler structural quantities. For all proteins in our study (Table~\ref{tab:pdb_names}), we used the Rosetta protein-design platform \citep{LeaverFayetal2011} to generate 500 designed variants. We then calculated the sequence entropy at each alignment position of the designed variants. We refer to the resulting quantity as the \emph{designed entropy}. We chose a design temperature of $T=0.6$, which was near the optimal range in our previous work \citep{Jacksonetal2013}.

\subsection*{Evaluating structural predictors of evolutionary sequence variation}

We began by comparing the Spearman correlations of sequence entropy with six different measures of local structural flexibility: B factors, RMSF obtained from MD simulations (MD RMSF), and RMSF obtained from crystal structures (CS RMSF), and variability in backbone and side-chain dihedral angles ($\phi$, $\psi$, and $\chi_1$). The correlation strengths of these quantities with entropy are shown in Figure~\ref{fig:cor_entropy_SF}. Significant correlations ($P<0.05$) are shown with filled symbols, and non-significant correlations are shown with empty symbols ($P\geq0.05$). We found that the variability in backbone dihedral angles, Var($\phi$) and Var($\psi$), explained the least variation in sequence entropy, while the variability in the side-chain dihedral angle, Var($\chi_1$), explained, on average, more variation in sequence entropy than did any other measure of structural flexibility. B factors and the two measures of RMSF explained, on average, approximately the same amount of variation in entropy, even though the results for individual proteins were somewhat discordant (see also next sub-section).

\begin{figure}[tbh]
\begin{center}
    \includegraphics[width=5in]{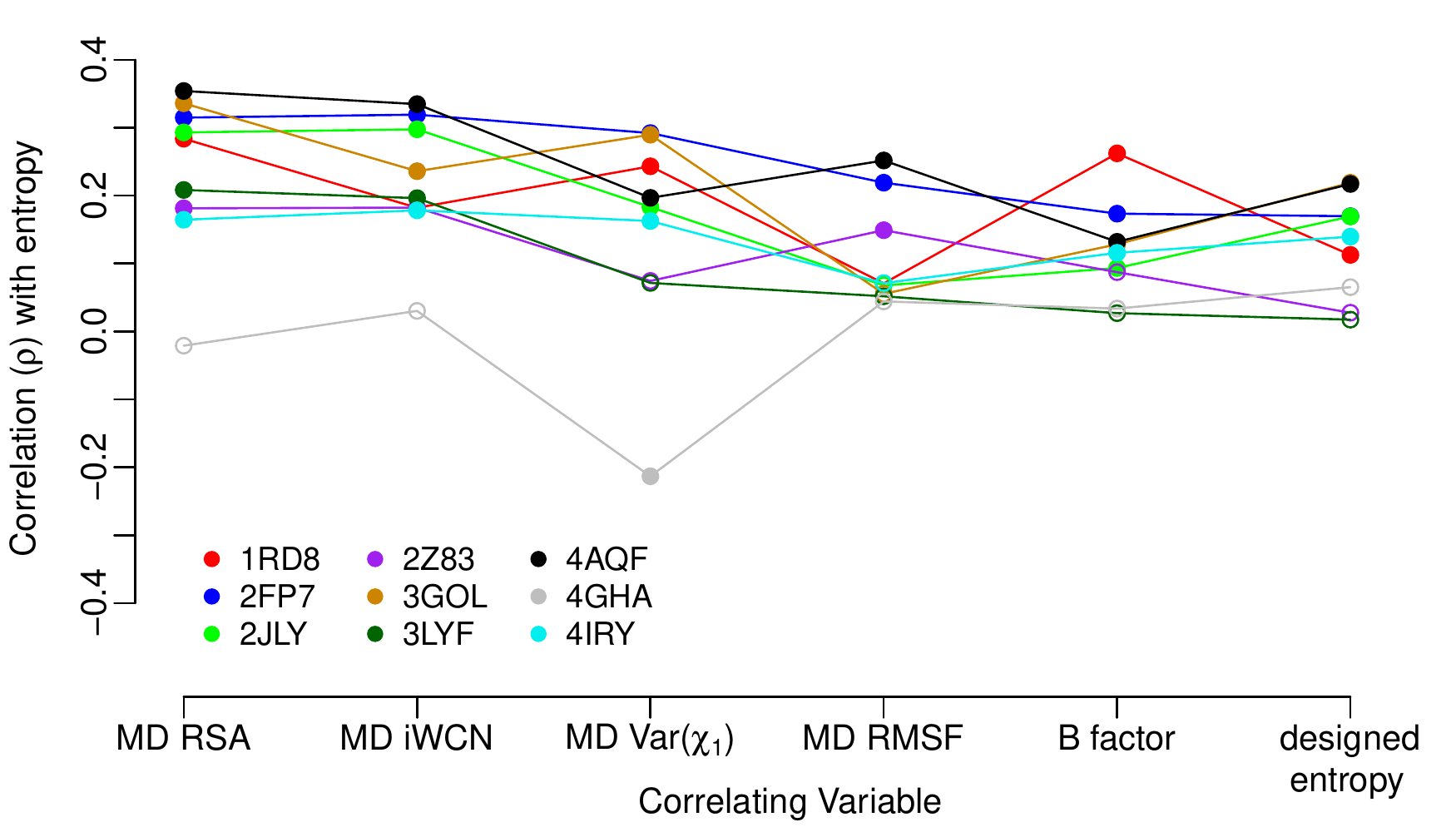} 
\end{center}
\caption{Spearman correlation of sequence entropy with measures of buriedness, packing density, and structural flexibility, as well as with designed entropy. Each symbol represents one correlation coefficient for one protein structure. Significant correlations ($P<0.05$) are shown as filled symbols, and insignificant correlations ($P\geq0.05$) are shown as open symbols. The quantities MD RSA, MD iWCN, MD Var($\chi_1$), and MD RMSF were calculated as time-averages over 15ns of MD simulations. B factors were obtained from crystal structures, and designed entropy was obtained from protein design in Rosetta. Compared to the measures of structural variability and to designed entropy, MD RSA and MD iWCN consistently show stronger correlations with sequence entropy. Note that results for MD iWCN are largely identical to those for MD iCN, so only MD iWCN was included here.}
\label{fig:cor_entropy_all}
\end{figure}

Based on results from the above analysis, we proceeded to compare the relative explanatory power among the best-performing measures of structural flexibility (Var$(\chi_1)$, MD RMSF, and B factors) with buriedness (RSA), packing density (iWCN), and designed entropy. Figure~\ref{fig:cor_entropy_all} shows the Spearman correlation coefficients between sequence entropy and each of the aforementioned quantities, for all proteins in our analysis. In this figure, several patterns emerged. First, nearly all correlations were positive and most were statistically significant, with the main exception of the Marburg virus RNA binding domain (PDB ID 4GHA). This protein only showed a single significant negative correlation between sequence entropy and Var($\chi_1$). Second, correlations were generally weak, such that no correlation coefficient exceeded 0.4. Third, on average, correlations were strongest for RSA and iWCN, yielding average correlations of $\rho=0.23$ and $\rho=0.22$, respectively. Fourth, designed entropy performed worse than RSA or iWCN as a predictor of evolutionary sequence variability, but it performed roughly the same as the three flexibility measures in this figure; the values of designed entropy, Var($\chi_1$), MD RMSF, and B factors showed average correlations of $\rho=0.13$, $\rho=0.14$, $\rho=0.11$, and $\rho=0.12$, respectively.

\subsection*{MD time-averages vs.\ crystal-structure snapshots}

Except for analyses involving B factors and CS RMSF, we obtained structural measures by averaging quantities over MD trajectories. This approach, however, did not reflect conventional practice for measuring RSA, CN, or WCN, which are typically measured from individual crystal structures. Therefore, we examined whether MD time-averages differed in any meaningful way from estimates obtained from crystal structures, and whether these estimates differed in their predictive power for evolutionary sequence variation.

\begin{figure}[tbh]
\begin{center}
    \includegraphics[width=6.5in]{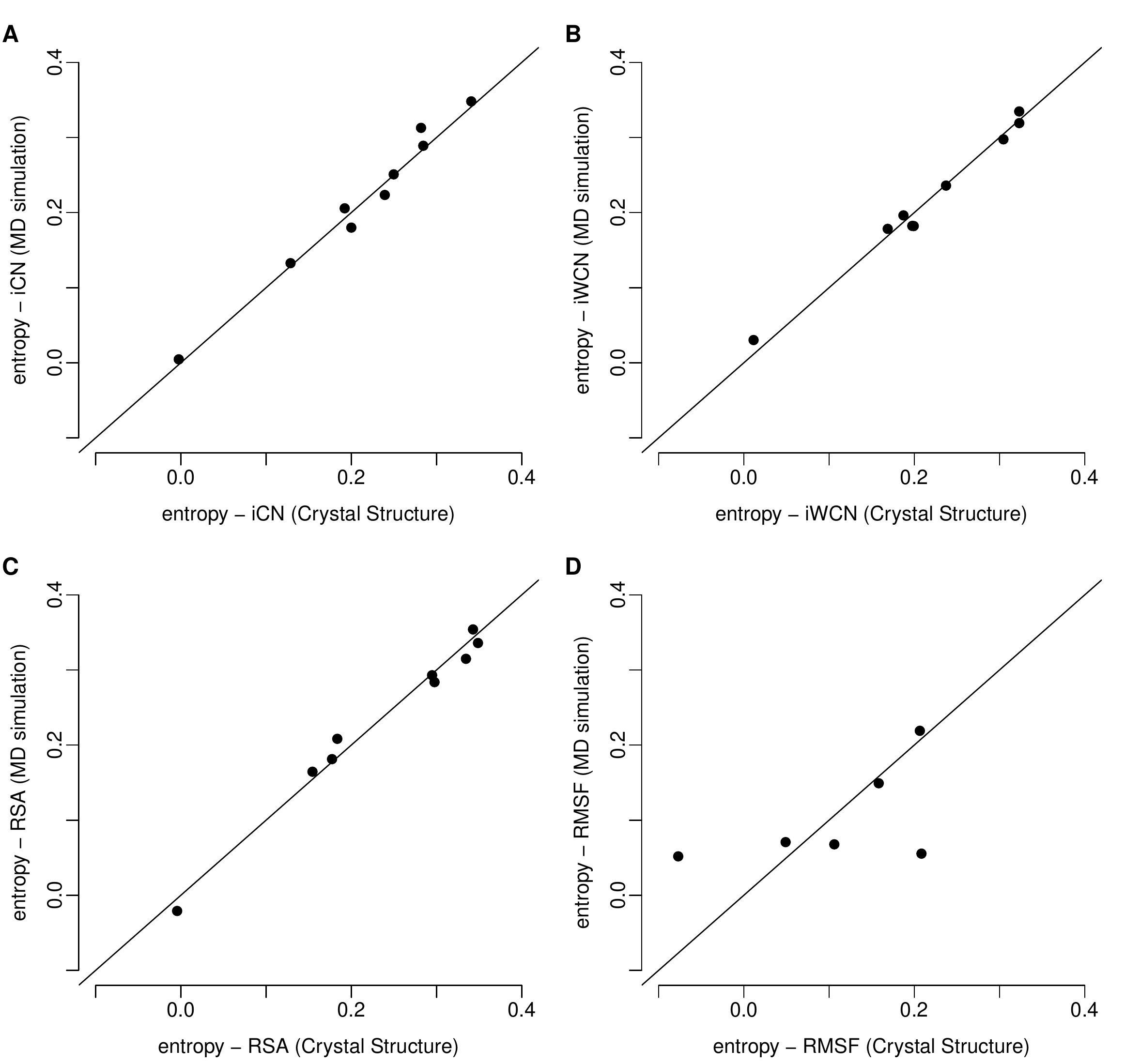} 
\end{center}
\caption{Spearman correlations of sequence entropy with MD-derived and crystal-structure derived structural measures. The vertical axes in all plots represent the Spearman correlation of sequence entropy with one structural variable obtained from $15ns$ of molecular dynamics (MD) simulations. The horizontal axes represent the Spearman's rank correlation coefficient of sequence entropy with the same structural variable as in the vertical axes but measured from protein crystal structures. Each dot represents one correlation coefficient for one protein structure. The quantities iCN, iWCN, and RSA have nearly identical predictive power for sequence entropy regardless of whether they are derived from MD simulations or from crystal structures. By contrast, MD RMSF yielded very different correlations than did CS RMSF.}
\label{fig:cor_cr_md}
\end{figure}

As shown in Table~\ref{tab:MDcrystal_cor}, RSA, CN, and WCN from crystal structures were highly correlated with their corresponding MD trajectory time-averages, for all protein structures we examined (Spearman correlation coefficients of $>0.9$ in all cases). Furthermore, the correlation coefficients we obtained when comparing the crystal-structure based measures to sequence entropy were virtually identical to coefficients obtained from the MC trajectory correlations (Figure~\ref{fig:cor_cr_md}A-C). Thus, in terms of predicting evolutionary variation, RSA, CN, and WCN values obtained from static structures performed as well as their MD equivalents averaged over short time scales. 

\begin{figure}[tbh]
\begin{center}
    \includegraphics[width=5in]{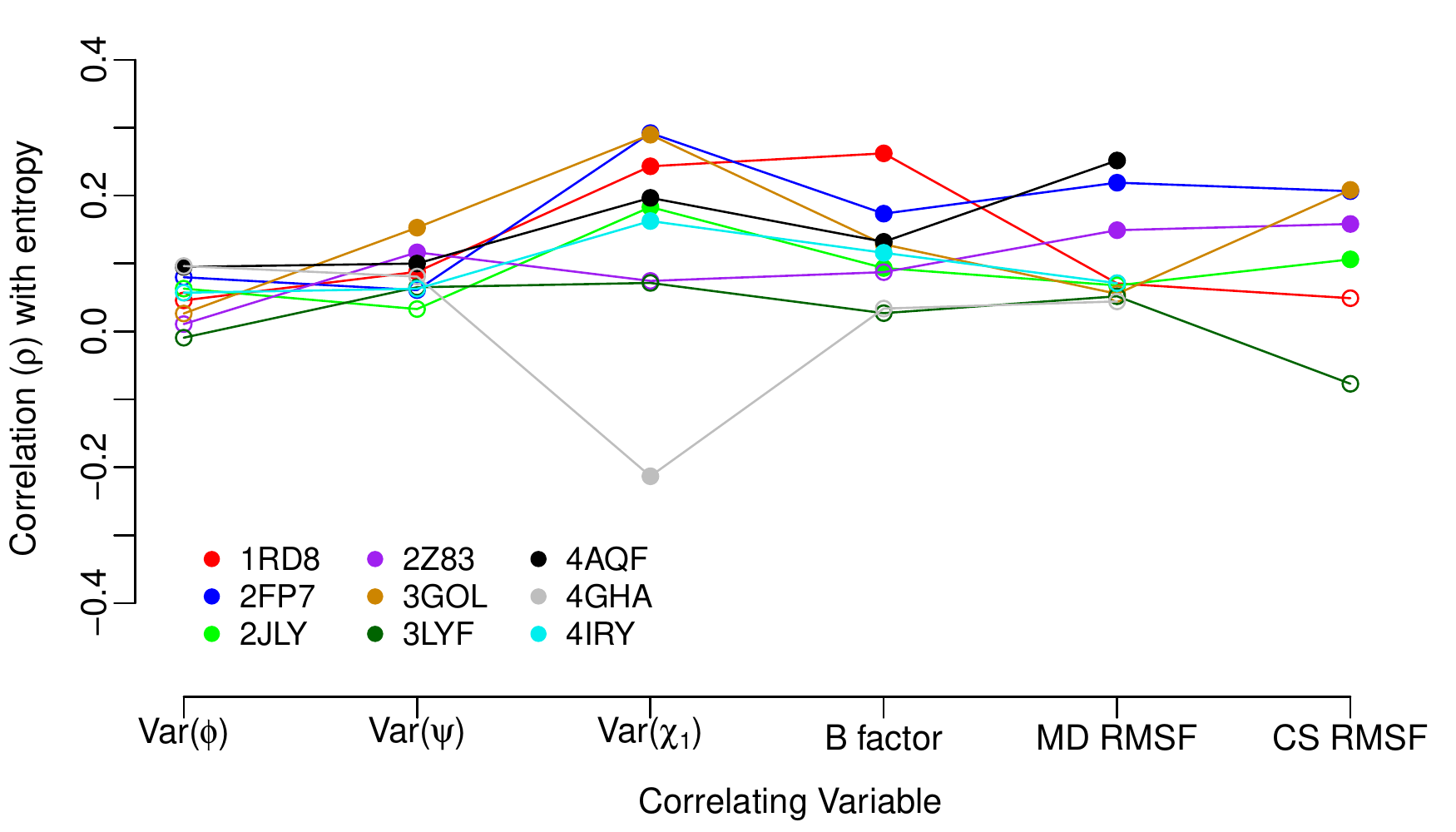} 
\end{center}
\caption{Spearman correlation of sequence entropy with measures of structural variability. Each symbol represents one correlation coefficient for one protein structure. Significant correlations ($P<0.05$) are shown as filled symbols, and insignificant correlations ($P\geq0.05$) are shown as open symbols. The quantities Var($\psi$), Var($\phi$), Var($\chi_1$), and MD RMSF were obtained as time-averages over 15ns of MD simulations. B factors were obtained from individual crystal structures. CS RMSF values were obtained from alignments of homologous crystal structures when available. Almost all structural measures of variability correlate weakly, but significantly, with sequence entropy.}
\label{fig:cor_entropy_SF}
\end{figure}

By contrast, correlations between corresponding MD RMSF to CS RMSF measures were sometimes quite different, with correlation coefficients ranging from 0.218 to 0.723 (Table~\ref{tab:MDcrystal_cor}). Consequently, for the two proteins for which MD RMSF was the least correlated with CS RMSF (hepatitis C protease and Rift Valley fever nucleoprotein), the strength of correlation between site entropy and RMSF depended substantially on how RMSF was calculated (Figures~\ref{fig:cor_entropy_SF} and~\ref{fig:cor_cr_md}D).

\begin{figure}[tbh]
\begin{center}
    \includegraphics[width=6.5in]{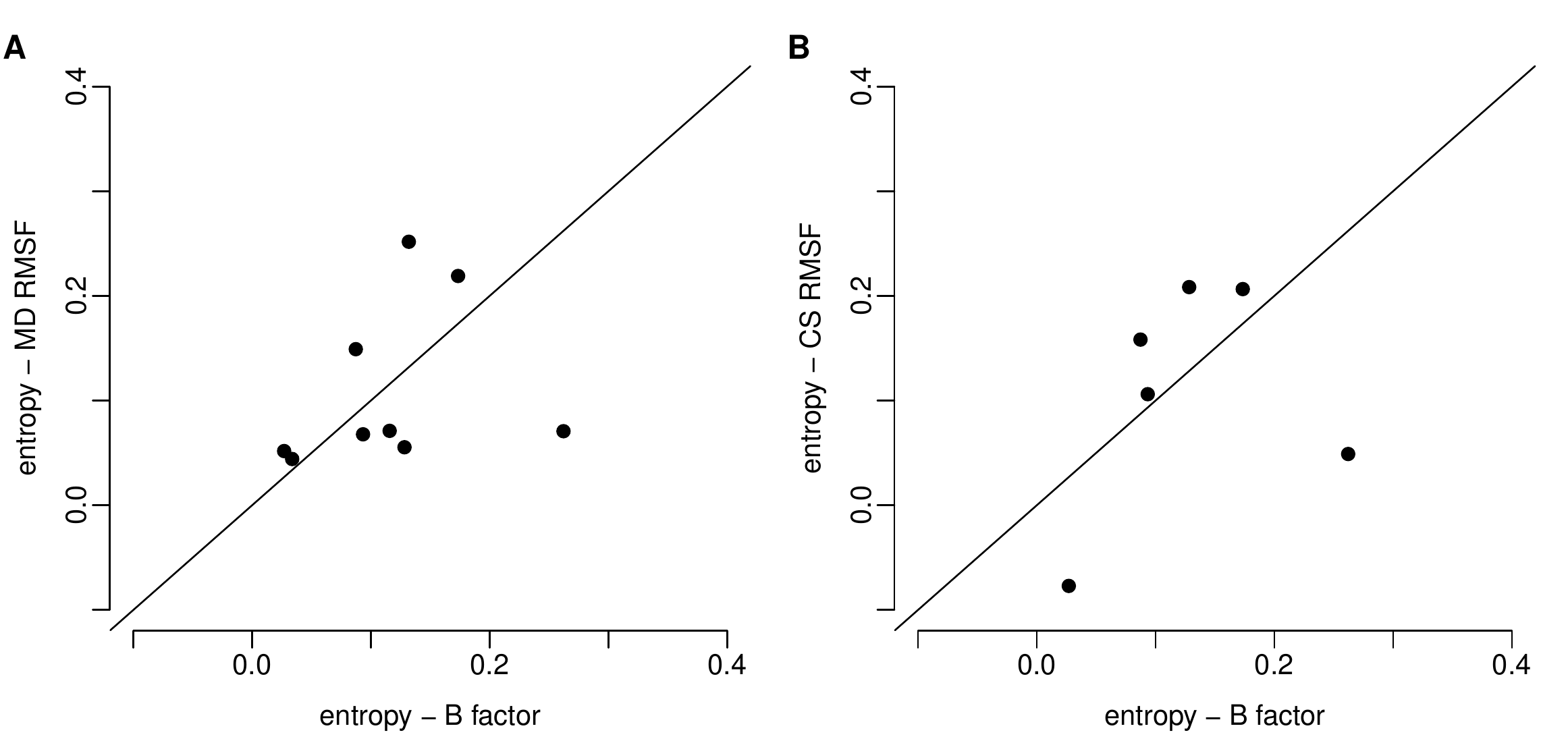} 
\end{center}
\caption{Spearman correlations of sequence entropy with measures of structural variability. Vertical and horizontal axes represent Spearman correlations of the indicated quantities. Each dot represents one correlation coefficient for one protein structure. MD RMSF, CS RMSF, and B factors all explain different amounts of variance in sequence entropy for different proteins.}
\label{fig:cor_entropy_bfca_rmsf}
\end{figure}

Finally, we examined whether correlations between sequence entropy and B factors or the two RMSF measures were comparable (Figure \ref{fig:cor_entropy_bfca_rmsf}). Again, we found that correlations between sequence entropy and B factors were generally different from those obtained for both MD RMSF and CS RMSF. This result highlighted that, while B factors, MD RSMF, and CS RMSF all measure backbone flexibility, they each contain distinct information about evolutionary sequence variability in our data set.

\subsection*{Sequence entropy vs.\ evolutionary-rate ratio $\omega$}

In the previous subsections, we used sequence entropy as a measure of site-wise evolutionary variation. While sequence entropy is a simple and straightforward measure of site variability, it has two potential drawbacks. First, while measured from homologous protein alignments, sequence entropy doesn't correct for the phylogenetic relationship of those alignment sequences. Hence, entropy can be biased if some parts of the phylogeny are more densely sampled than others. Second, entropy does not take the actual substitution process into account. As a result, a single substitution near the root of the tree can result in a comparable entropy to a sequence of substitutions toggling back and forth between two amino acids.

\begin{figure}[tbh]
\begin{center}
    \includegraphics[width=4in]{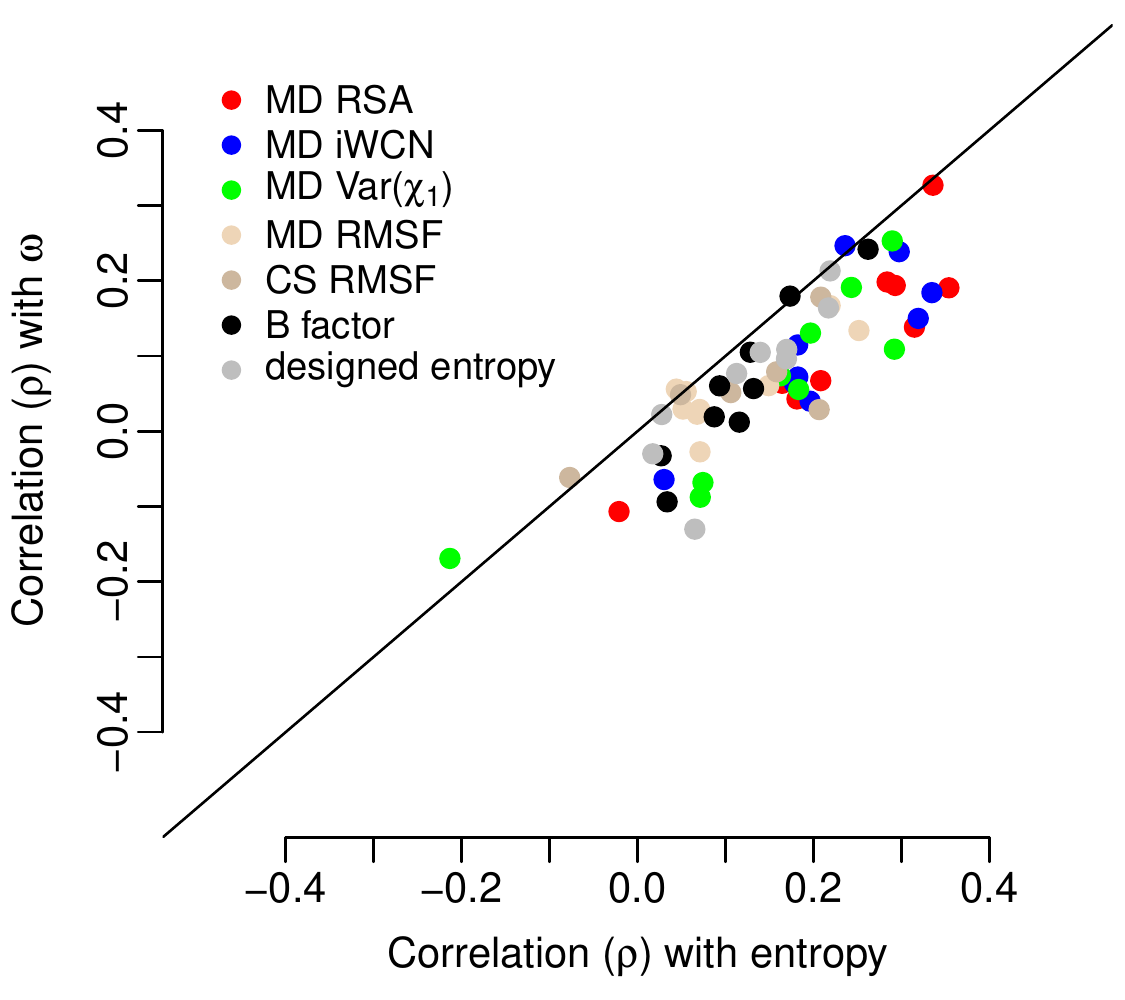} 
\end{center}
\caption{Spearman correlations of structural quantities with sequence entropy and with the evolutionary rate ratio $\omega$. Nearly all points fall below the $x=y$ line, indicating that structural quantities generally predict as much as or more variation in sequence entropy than in $\omega$.}
\label{fig:cor_entropy_omega}
\end{figure}

To consider an alternative quantity of evolutionary variation that doesn't suffer from either of these drawbacks, we calculated the evolutionary-rate ratio $\omega=dN/dS$ for all proteins at all sites, and repeated all analyses with $\omega$ instead of entropy. We found that results generally carried over, but with somewhat weaker correlations. Figure~\ref{fig:cor_entropy_omega} plots, for each protein, the Spearman correlations between $\omega$ and our various predictors versus the correlation between entropy and our predictors. Most data points fall below the $x=y$ line and are shifted downwards by approximately 0.1. Thus, correlations of structural quantities and designed entropy with $\omega$ are, on average, approximately 0.1 smaller than correlations of the same quantities with sequence entropy.

\subsection*{Multi-variate analysis of structural predictors}

The various structural quantities we have considered are by no means independent of each other. Measures of buriedness and packing density co-vary with each other, as do measures of structural flexibility. Further, the latter co-vary with the former, as does designed entropy. Therefore, we conducted a joint multivariate analysis, which included most structural quantities considered in this work. We employed this strategy to determine the extent to which these quantities contained independent information about sequence variability while additionally
assessing whether combining multiple structural quantities yielded improved predictive power. We employed a principal component (PC) regression approach, which has previously been used successfully to disentangle genomic predictors of whole-protein evolutionary rates \citep{Drummondetal2006,Bloometal2006}. For each analysis described below, we first carried out a PC analysis of the predictor variables (i.e., the structural quantities such as RSA and RMSF), and we subsequently regressed the response (either sequence entropy or $\omega$) against the individual components. Note that variables were not rank-transformed for this analysis.

For a first PC analysis, we pooled all structural quantities and then regressed entropy against each PC separately, for all proteins in our data set. This strategy allowed us to analyze all proteins in our data set individually but in such a way that results were comparable from one protein to the next. We excluded CS RMSF from this analysis, so that we could include results from all nine viral proteins. The results of this analysis are shown in Figure~\ref{fig:cor_entropy_PC_screen}. The first component (PC1) explained, on average, the largest amount of variation in sequence entropy (see Figure~\ref{fig:cor_entropy_PC_screen}A). PC3 yielded the second-highest $r^2$ value, on average, while all other components explained very little variation in sequence entropy. When looking at the composition of the components, we found that RSA, iWCN, RMSF, and Var($\chi_1$) all loaded strongly on PC1, while PC2 and PC3 where primarily represented by designed entropy and B factors (see Figure~\ref{fig:cor_entropy_PC_screen}B and C). RMSF also had moderate loadings on PC3. Interestingly, designed entropy and B factors load with equal signs on PC2 but with opposite signs on PC3.

We interpreted PC1 to represent a buriedness/packing-density component. By definition, PC1 measures the largest amount of variation among the structural quantities, and all structural quantities reflect to some extent the buriedness of residues and the number of residue-residue contacts. PC2 and PC3 were more difficult to interpret. Since designed entropy and B factors loaded strongly on both but with two different combinations of signs, we concluded that the most parsimonious interpretation was to consider PC2 as a component representing sites with high designed entropy and high spatial fluctuations (as measured by B factors) and PC3 representing sites with high designed entropy and low spatial fluctuations. Using these interpretations, our PC regression analysis suggested that of all the structural quantities considered here, residue buriedness/packing was the best predictor of evolutionary variation. Designed entropy was a useful predictor as well, but it tended to perform better at sites with low spatial fluctuations.

For a second PC analysis, we included the predictor CS RMSF, which therefore restricted the data set to include only six proteins (see Table~\ref{tab:homologs}). This analysis, which retained sequence entropy as the response variable, yielded comparable results to the first PC analysis. The main differences occurred in PC2 and PC3, where CS RMSF generally loaded in the opposite direction of B factor, and either in the same (PC2) or the opposite (PC3) direction of designed entropy (Figure~\ref{fig:cor_entropy_PC_screen_CSrmsf}). 

Finally, we redid the two PC analyses described above, but instead with $\omega$ as the response variable (Figures~\ref{fig:cor_omega_PC_screen} and~\ref{fig:cor_omega_PC_screen_CSrmsf}). Again, these results were largely comparable to results from PC analyses with sequence entropy as the response.

\section*{Discussion}

We have carried out a comprehensive analysis of the extent to which different structural quantities predict sequence evolutionary variation in nine viral proteins. We found that measures of buriedness and local packing generally performed better than did measures of structural flexibility. Further, the former measures also performed better than a computational protein-design approach that employed a sophisticated all-atom force field to determine allowed amino-acid distributions at each site. Finally, there was no difference in predictive power between structural quantities obtained from averaging structural quantities over 15ns of MD simulations versus taking the same quantities from individual crystal structures. 

Our results are broadly in agreement with recent work by Echave and collaborators \citep{Yehetal2014,Huangetal2014}. These authors found that RSA and CN showed comparable correlation strengths with evolutionary sequence variation \citep{Yehetal2014}. Further, they demonstrated that the observed relationship between evolutionary variation and residue--residue contacts was not consistent with a flexibility model that puts evolutionary variability in proportion to structural flexibility \citep{Huangetal2014}. Instead, a mechanistic stress model, in which amino-acid substitutions cause physical stress in proportion to the number of residue--residue contacts affected, could explain all the observed data \citep{Huangetal2014}.

The correlation strengths we observed were consistently lower than those observed previously \citep{Jacksonetal2013,Yehetal2014}. We believe that this result was due to our choice of analyzing viral proteins instead of the cellular proteins or enzymes used in prior works. First, while viral sequences are abundant, their alignments may not be as diverged as alignments that can be obtained for sequences from cellular organisms. For example, our influenza sequences spanned only approximately one decade. Despite the high mutation rates observed in RNA viruses, the evolutionary variation that can accumulate over this time span is limited. This relatively lower evolutionary divergence makes resolving differences between more and less conserved sites much more difficult. Second, many viral proteins experience a substantial amount of selection pressure to evade host immune responses. The resulting positive selection on viral sequences may mask evolutionary constraints imposed by structure. For example, influenza hemagglutinin displays positive selection throughout the entire sequence, regardless of the extent of residue burial \citep{MeyerWilke2013,Meyeretal2013,Suzuki2006,Bushetal1999}. However, the results we obtained here for viral proteins are broadly consistent with the results obtained earlier for cellular proteins \citep{DokholyanShakhnovich2001,FranzosaXia2009,Jacksonetal2013,Yehetal2014}, indicating that viral proteins evolve under many of the same biophysical selection pressures that cellular proteins experience.

We have found here that correlations between sequence entropy and structural quantities were consistently higher than correlations between the evolutionary-rate ratio $\omega$ and structural quantities. Surprisingly, in a recent study on cellular proteins, \citet{Yehetal2014b} found that entropy performed worse than quantities assessing substitution rates. One possible explanation for this discrepancy is again our choice of viral sequences. Our sequence alignments almost certainly contained some polymorphisms, whereas the sequences of \citet{Yehetal2014b} likely did not. It is known that polymorphisms may diminish the reliability of $\omega$ estimates \citep{KryazhimskiyPlotkin2008}. While the effect of polymorphisms on sequence entropy is not known, it seems plausible that entropy would be less sensitive to them than $\omega$ is. Alternatively, since viral proteins frequently experience positive selection, rate estimates may be confounded by this selection pressure and thus less reflective of constraints imposed by protein structure. By contrast, even under positive selection amino-acid distributions at sites would have to be consistent with the constraints imposed by the protein structure, and entropy would remain sensitive to these constraints.

We found that simple measures of buriedness or packing density, such as RSA or CN, were better predictors of evolutionary variation than was sequence variability predicted from computational protein design. In other words, simple quantities that can be obtained trivially from PDB structures performed better than a sophisticated protein-design strategy that makes use of an all-atom energy function and requires thousands of CPU-hours to complete. This result highlights that, even though computational protein design has yielded impressive results in specific cases \citep{Kuhlman2003,Rothlisberger2008,Fleishman2011}, this approach remains limited in its ability to predict evolutionary variation. Similarly, we have previously found that flexible backbone design with Rosetta produced designs whose surface and core were too similar \citep{Jacksonetal2013}. We attributed this discrepancy to either the solvation model or the model of backbone flexibility we used (Backrub, see \citealt{Smith2008}). The results we found here suggest that the model of backbone flexibility may indeed be the cause of at least some of the discrepancies between predicted and observed site variability. In particular, in our PC regression analysis, the component in which designed entropy loaded opposite to B factor and MD RMSF generally had the second-highest predictive power for evolutionary variability, after the component representing buriedness/packing density. In sum, designed entropy was a better predictor for evolutionary sequence variability for sites with less structural flexibility compared to sites with more flexibility.

Even though RSA and CN remain the best currently known predictors of evolutionary variation, neither quantity has particularly high predictive power. One reason why predictive power may be low is that neither quantity accounts for correlated substitutions at interacting sites.  Yet such correlated substitutions happen regularly. For example, covariation among sites encodes information about residue-residue contacts and 3D structure \citep{Halabietal2009,BurgervanNimwegen2010,Marksetal2011,Jonesetal2014}, and evolutionary models that incorporate residue--residue interactions tend to perform better than models that do not \citep{Rodrigueetal2005,BordnerMittelmann2014}. An improved predictor of evolutionary variation would have to correctly predict this covariation from structure. In principle, computational protein design, which takes into consideration the atom-level details of the protein structure, should properly reproduce covariation among sites. However, a recent analysis showed that there are significant limitations to the covariation that is predicted \citep{OllikainenKortemme2013}. In addition, covariation in designed proteins is quite sensitive to the type of backbone variation modeled during design, and improved models of backbone flexibility may be required for improved prediction of covariation among sites \citep{OllikainenKortemme2013}.

\section*{Acknowledgments}

This work was supported in part by NIH grant R01 GM088344, DTRA grant HDTRA1-12-C-0007, ARO grant W911NF-12-1-0390, and the BEACON Center for the Study of Evolution in Action (NSF Cooperative Agreement DBI-0939454). The Texas Advanced Computing Center at UT Austin provided high-performance computing resources.


\newpage

\section*{Tables}

\begin{table}[htbp]
\caption{PDB structures considered in this study.\label{tab:pdb_names}}
\smallskip

\centerline{
\begin{tabular}{c c c c c} 
                        \hline
    Viral Protein   &  PDB ID  & Chain    & Sequence  & Number of   \\ 
    &    &       & Length    & Sequences   \\\hline                                                           
  Hemagglutinin Precursor         & 1RD8           & AB        & 503                & 1039        \\
  Dengue Protease Helicase        & 2JLY           & A         & 451                & 2362        \\
  West Nile Protease              & 2FP7           & B         & 147                & 237         \\
  Japanese Encephalitis Helicase  & 2Z83           & A         & 426                & 145         \\
  Hepatitis C Protease            & 3GOL           & A         & 557                & 1021        \\
  Rift Valley Fever Nucleoprotein & 3LYF           & A         & 244                & 95          \\
  Crimean Congo Nucleocapsid      & 4AQF           & B         & 474                & 69          \\
  Marburg RNA Binding Domain      & 4GHA           & A         & 122                & 42          \\
  Influenza Nucleoprotein         & 4IRY           & A         & 404                & 943         \\
  \hline
\end{tabular}
}
\end{table}

\begin{table}[htbp]
\caption{Availability of homologous crystal structures. Although most viral proteins have many PDB structures available, the sequence divergence among these structures is low. Therefore, when calculating RMSF from crystal structures, we considered only those proteins with at least five homologous structures at 5\% pairwise sequence divergence (highlighted in bold). \label{tab:homologs}}
\smallskip

\centerline{
\begin{tabular}{c c c c c c}
                        \hline
  Viral Protein & BLAST hits$^\text{a}$ & \multicolumn{4}{c}{Unique sequences} \\ 
    \cline{3-6}\noalign{\smallskip}
                &            & all  & $\geq2$\%$^\text{b}$  &  $\geq5$\%$^\text{b}$  & $\geq10$\%$^\text{b}$ \\\hline
  Hemagglutinin Precursor         & 63 & 17 & 10 & \textbf{9} & 7\\
  Dengue Protease Helicase        & 31 & 13 & 7 & \textbf{7} & 7\\
  West Nile Protease              & 21 & 16 & 10 & \textbf{7} & 6\\
  Japanese Encephalitis Helicase  & 31 & 12 & 7 & \textbf{7} & 7\\
  Hepatitis C Protease            & 302 & 33 & 10 & \textbf{5} & 4 \\
  Rift Valley Fever Nucleoprotein & 95  & 9 & 5 & \textbf{5} & 5\\
  Crimean Congo Nucleocapsid      & 7 & 4 & 3 & 2 & 2\\
  Marburg RNA Binding Domain      & 63 & 9 & 5 & 3 & 3\\
  Influenza Nucleoprotein         & 69 & 15 & 4 & 4 & 2\\
  \hline
\end{tabular}
}
$^\text{a}$ BLAST hits against all sequences in the PDB, excluding hits with $<35$\% sequence identity and $<90$\% alignment length\\
$^\text{b}$ Unique sequences at indicated minimum pairwise sequence divergence
\end{table}

\begin{table}[htbp]
\caption{Correlations between quantities obtained from MD trajectories and from crystal structures. For each quantity and each protein, we calculated the Spearman correlation $\rho$ between the values obtained from MD time averages and the values obtained from viral protein crystal structures. Note that crystal structures for all nine proteins were used for RSA, CN, and WCN calculations, but only the six proteins for which we had sufficient crystal structure variability were used for CS RMSF. We then calculated the minimum, maximum, mean, and standard deviation of these correlations.\label{tab:MDcrystal_cor}}
\bigskip

\centerline{
\begin{tabular}{ccccc}
                        \hline
  Quantity    & min $\rho$ & max $\rho$ & $\langle\rho\rangle$ & SD($\rho$) \\\hline
  RSA       & 0.937 & 0.981 & 0.948 & 0.012\\
  CN        & 0.964 & 0.993 & 0.976 & 0.008\\
  WCN       & 0.973 & 0.991 & 0.984 & 0.006\\
  RMSF      & 0.218 & 0.723 & 0.502 & 0.181\\
  \hline
\end{tabular}
}

\end{table}

\clearpage

\section*{Figures}

\begin{figure}[tbh]
\begin{center}
      \includegraphics[width=6.5in]{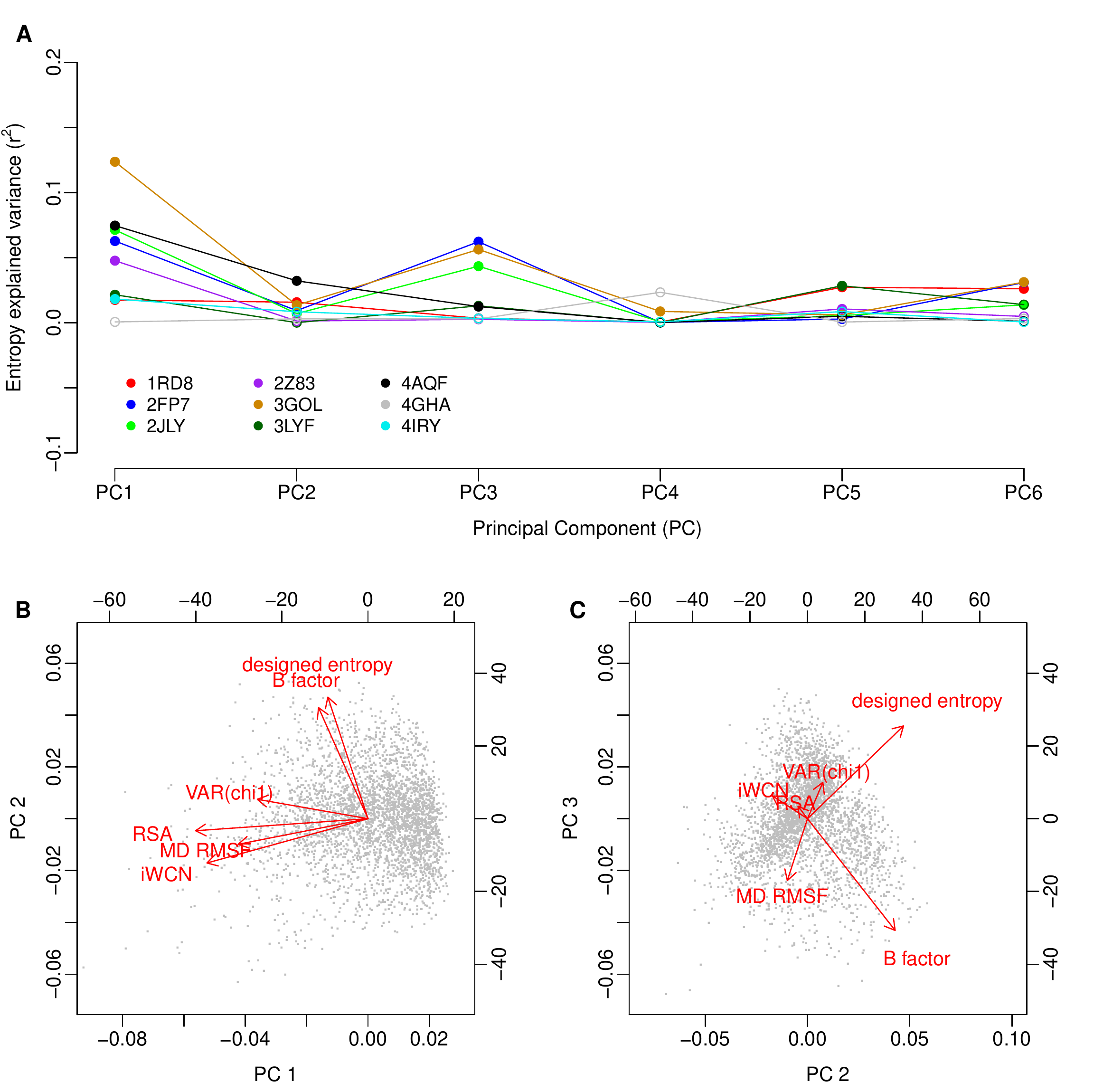} 
\end{center}
\caption{Principal Component (PC) Regression of sequence entropy against the structural variables. {\bf (A)} Variance in entropy explained by each principal component. For most proteins, PC1 and PC3 show the strongest correlations with sequence entropy. Significant correlations ($P<0.05$) are shown as filled symbols, and insignificant correlations ($P\geq0.05$) are shown as open symbols. {\bf (B)} and {\bf (C)} Composition of the three leading components. Red arrows represent the loadings of each of the structural variables on the principal components; black dots represent the amino acid sites in the PC coordinate system. The variables RSA, iWCN, MD RMSF, and Var($\chi_1$) load strongly on PC1 and weakly on PC2, while B factor and designed entropy load strongly on PC2 and weakly on PC1. Interestingly, B factor and designed entropy also load strongly on PC3, but in opposite directions.}
\label{fig:cor_entropy_PC_screen}
\end{figure}


\clearpage

\newpage
\section*{Supplementary Figures}

\centerline{\includegraphics[width=6.5in]{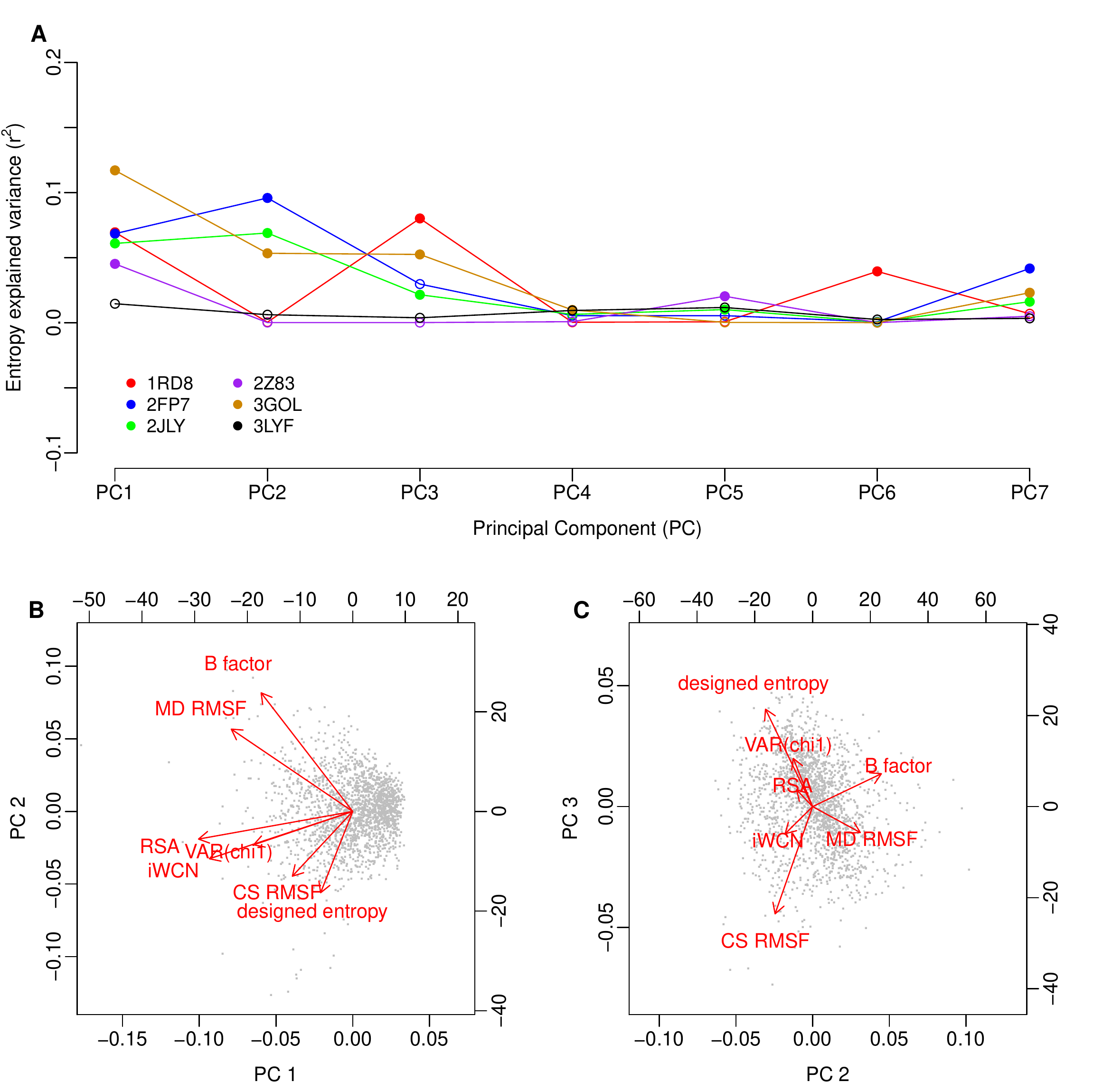}}
\noindent \textbf{Fig. S1} Principal Component (PC) Regression of sequence entropy against the structural variables, including CS RMSF. {\bf (A)} Variance in entropy explained by each principal component. For most proteins, PC1 and either PC2 or PC3 show the strongest correlations with sequence entropy. Significant correlations ($P<0.05$) are shown as filled symbols, and insignificant correlations ($P\geq0.05$) are shown as open symbols. {\bf (B)} and {\bf (C)} Composition of the three leading components. Red arrows represent the loadings of each of the structural variables on the principal components; black dots represent the amino acid sites in the PC coordinate system.
\customlabel{fig:cor_entropy_PC_screen_CSrmsf}{S1}

\newpage
\centerline{\includegraphics[width=6.5in]{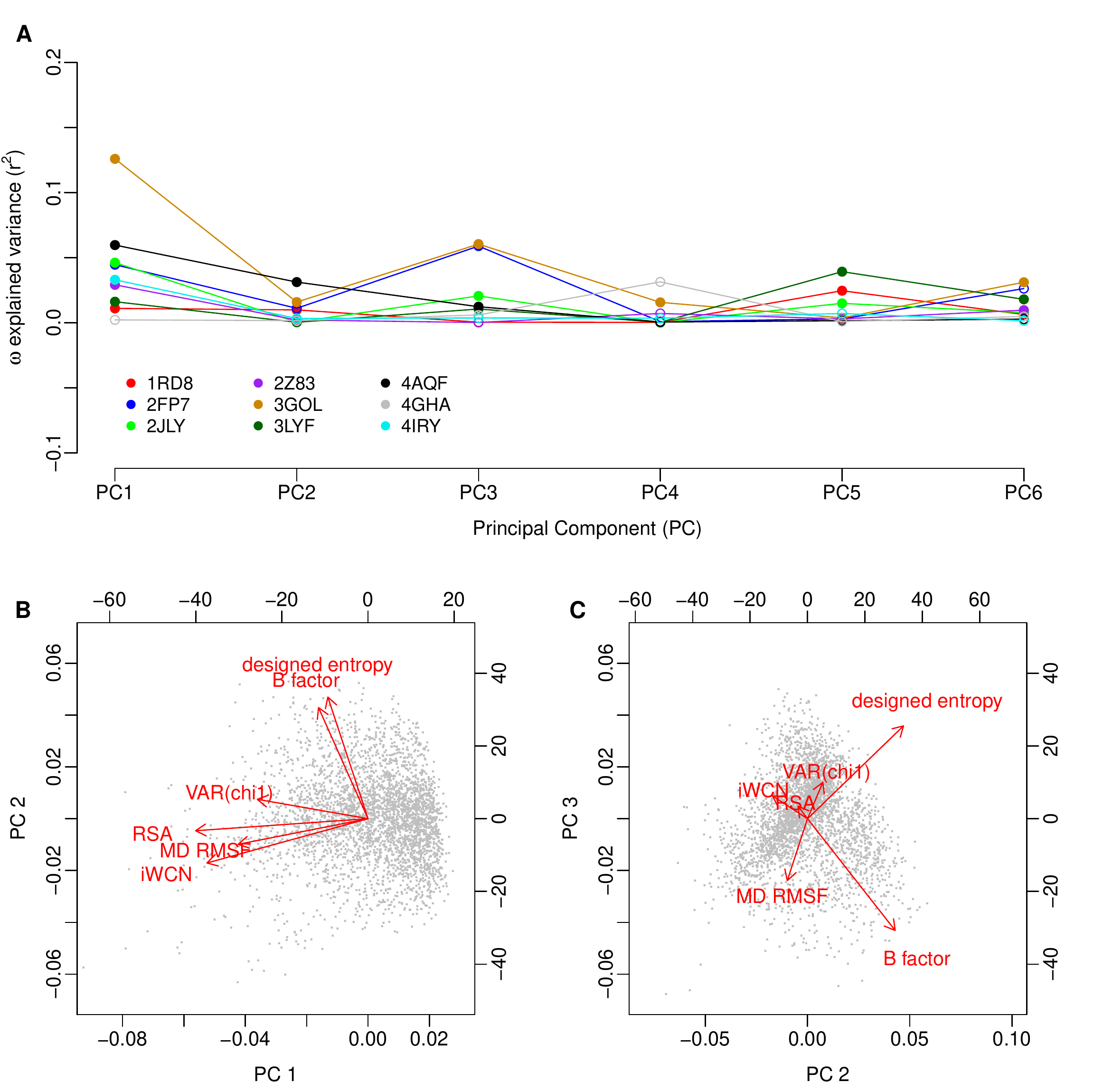}}
\noindent \textbf{Fig. S2} Principal Component (PC) Regression of $\omega$ against the structural variables. {\bf (A)} Variance in $\omega$ explained by each principal component. For most proteins, PC1 and PC3 show the strongest correlations with $\omega$. Significant correlations ($P<0.05$) are shown as filled symbols, and insignificant correlations ($P\geq0.05$) are shown as open symbols. {\bf (B)} and {\bf (C)} Composition of the three leading components. Red arrows represent the loadings of each of the structural variables on the principal components; black dots represent the amino acid sites in the PC coordinate system. Note that parts B and C are identical to those shown in Figure~\ref{fig:cor_entropy_PC_screen}.
\customlabel{fig:cor_omega_PC_screen}{S2}

\newpage
\centerline{\includegraphics[width=6.5in]{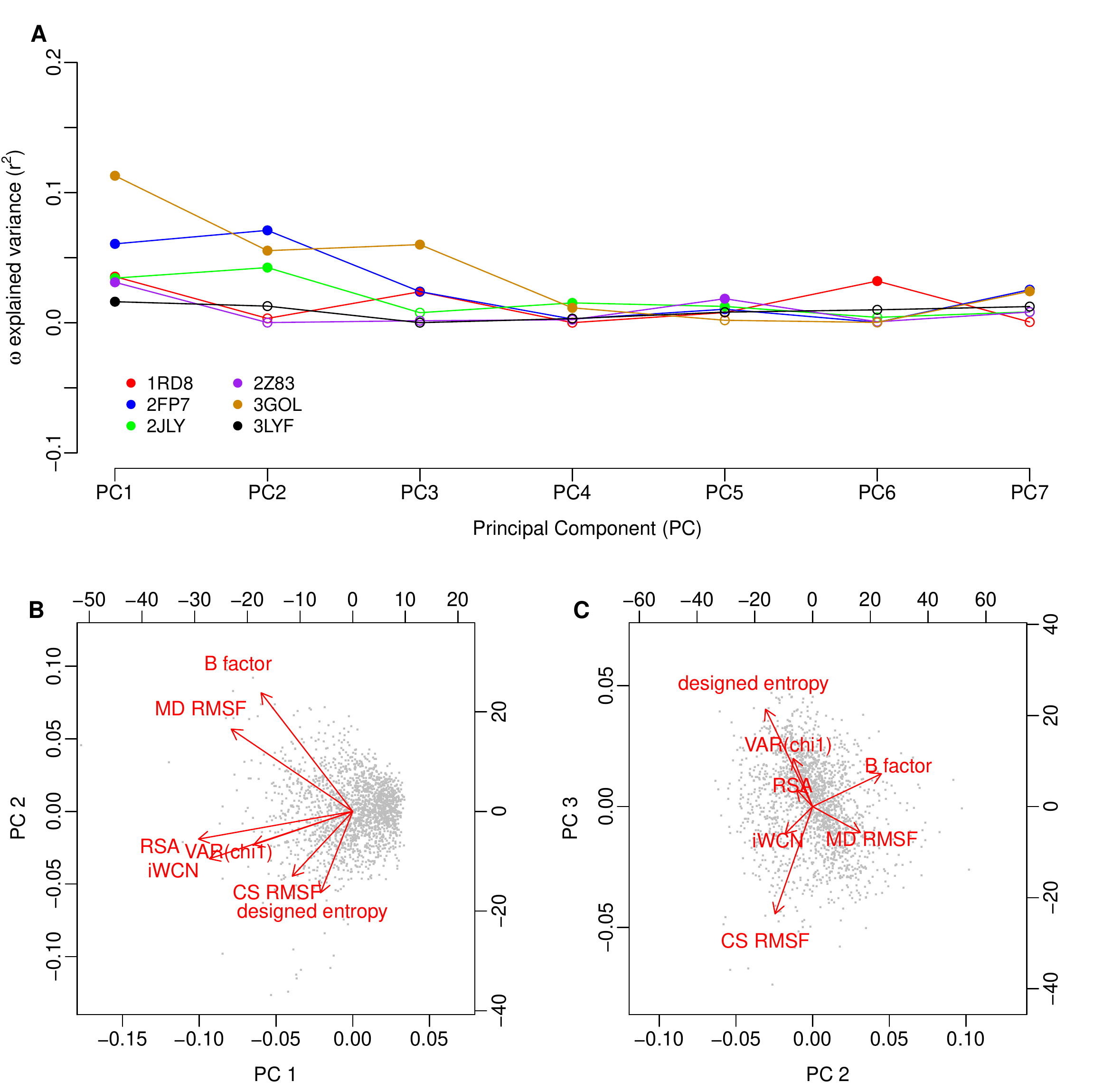}}
\noindent \textbf{Fig. S3} Principal Component (PC) Regression of $\omega$ against the structural variables, including CS RMSF. {\bf (A)} Variance in $\omega$ explained by each principal component. For most proteins, PC1 and either PC2 or PC3 show the strongest correlations with $\omega$. Significant correlations ($P<0.05$) are shown as filled symbols, and insignificant correlations ($P\geq0.05$) are shown as open symbols. {\bf (B)} and {\bf (C)} Composition of the three leading components. Red arrows represent the loadings of each of the structural variables on the principal components; black dots represent the amino acid sites in the PC coordinate system. Note that parts B and C are identical to those shown in Figure~\ref{fig:cor_entropy_PC_screen_CSrmsf}.
\customlabel{fig:cor_omega_PC_screen_CSrmsf}{S3}

\end{document}